\documentclass[11pt,twoside]{article}
\usepackage{asp2004}
\usepackage{psfig}
\usepackage{epsf}
\usepackage{graphics}
\usepackage{lscape}
\def\edcomment#1{\iffalse\marginpar{\raggedright\sl#1\/}\else\relax\fi}
\reversemarginpar

\begin{document}
\title{Extragalactic Synchrotron Sources at Low Frequencies}
\author{J. A. Eilek$^1$, F. N. Owen$^2$, and  T. Markovi\'c$^{1,2}$}
\affil{$^1$New  Mexico Tech, Socorro NM 87801 USA}
\affil{$^2$National Radio Astronomy Observatory, Socorro NM 87801 USA}

\begin{abstract}

The LWA will be well suited to address many important questions 
about the physics and astrophysics of extragalactic synchrotron 
sources.  Good low-frequency data will enable major steps forward
in our understanding of radio galaxy physics, of the plasma in
clusters of galaxies, and of active objects in  the high-redshift
universe.  Such data will also be important in answering some basic
questions about the physics of synchrotron-emitting plasmas. 

\end{abstract}
\thispagestyle{plain}

\section{General Introduction}

Synchrotron emission is one of our most important tools for understanding
the active extragalactic universe.  Energetic processes in many
astrophysical settings create a relativistic, magnetized plasma which
we can observe {\it via} its synchrotron emission.  We can study  the
high-energy activity directly, for instance in the creation of a radio  
jet.  We can use radio sources as probes of larger questions, for instance 
learning about the plasma atmosphere surrounding a radio tail by studying
the dynamics of the radio tail, or  understanding the dynamical state of the
dark matter in a cluster of galaxies by studying the synchrotron halo in
the cluster.  We can study the population of massive black holes in
the early universe through detections of high-redshift synchrotron
sources.

High quality images at low frequencies will be critical to all of
these applications.  In this paper we discuss some important
contributions the LWA will make.  One contribution is simple
exploration.  Many interesting sources are steep spectrum, and
therefore can best  be detected and studied at low
frequencies. Multi-frequency imaging will be another important
contribution, combining high-resolution images from the LWA with those
from  higher-frequency  instruments such as the EVLA.   We  need good
images  across a broad frequency range, including the very low
frequencies which the LWA will make available, in order to understand
the nature and dynamics of radio jets on large and small scales.
Finally, some fundamental questions regarding particle acceleration
and the physics of synchrotron emission are still unresolved;   good 
low-frequency data will be an important part of the answers.

\section{Basic Synchrotron Physics}

If we want to use synchrotron emission to study  active
extragalactic sources, we need to understand the physics of the emitting 
plasma.  Two important issues (at least) remain obscure, the low-frequency
``injection'' spectrum and the high-frequency turndown known as
 ``spectral aging''.  
We remind the reader that the electron energy, $\gamma m_e c^2$, maps
to the emission frequency in synchrotron radiation 
as $\nu_{sy} \propto \gamma^2 B$.  Thus, for a given magnetic field,
lower emission frequencies correspond to lower electron energies.

\subsection{Low Frequency Spectra}

We still do not know  how relativistic electrons are accelerated in
radio sources.  Options include shock acceleration, turbulent acceleration,
and electrodynamic acceleration (such as in a reconnection layer).  These
different mechanisms  have different signatures in the energy range
 and spectrum of the accelerated
electrons.  In particular, whether or not the low-energy electron spectrum 
is a power law is often assumed but has not been established from the data.
Some authors have disputed this assumption based on the data
(Katz-Stone {\it et al.\,}1993), and simple models of acceleration can lead to 
alternatives ({\it e.g.}, Borovsky \& Eilek 1986, Brunetti {\it et
  al.\,}2001).  The particle distribution at low 
energies is also important to the energy budget of the system; 
small differences in the usual assumptions (power laws and cutoffs)
can have large consequences on the interpretation ({\it e.g.}, Harris 2005).

Because radiative losses affect the highest energy electrons 
first, the low frequency synchrotron spectrum, coming from the 
low energy electrons, should be the most ``pristine''.  If we can 
measure the low-frequency spectrum  {\it in a homogeneous region}, we
can  determine both the spectrum and low-energy cutoff of the 
electrons.  This will allow critical tests of the different acceleration
models.   We note that the only unambiguous sign of a low-energy
cutoff in the electron distribution is the characteristic  $\nu^{1/3}$
low-frequency spectrum;  steeper low-frequency spectra are due to
absorption effects.  

\subsection{Spectral Aging}

Another important question is the interpretation of
 high-frequency spectral steepening.   This common effect 
is illustrated by 3C465 in Figure 1, which also shows how different a radio
source can look at low and high frequencies.   Other striking
examples include the lobes of Cyg A (Carilli {\it et al.\,}1991) and the long
tails recently detected in Hyd A (Lane {\it et al.\,}2004).

\begin{figure}[ht]

\begin{center}
\scalebox{0.224}{\includegraphics{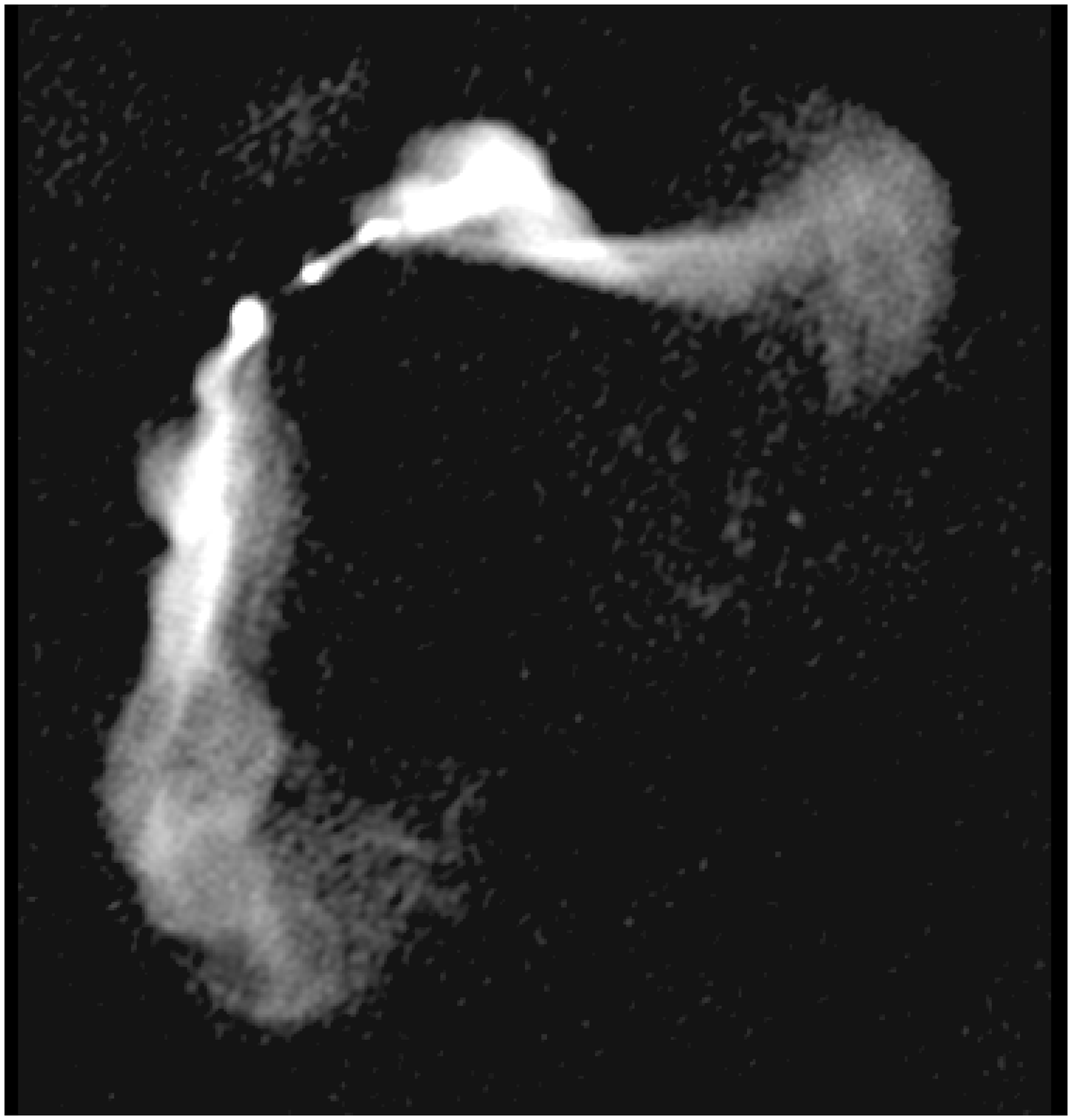}}
\scalebox{0.224}{\includegraphics{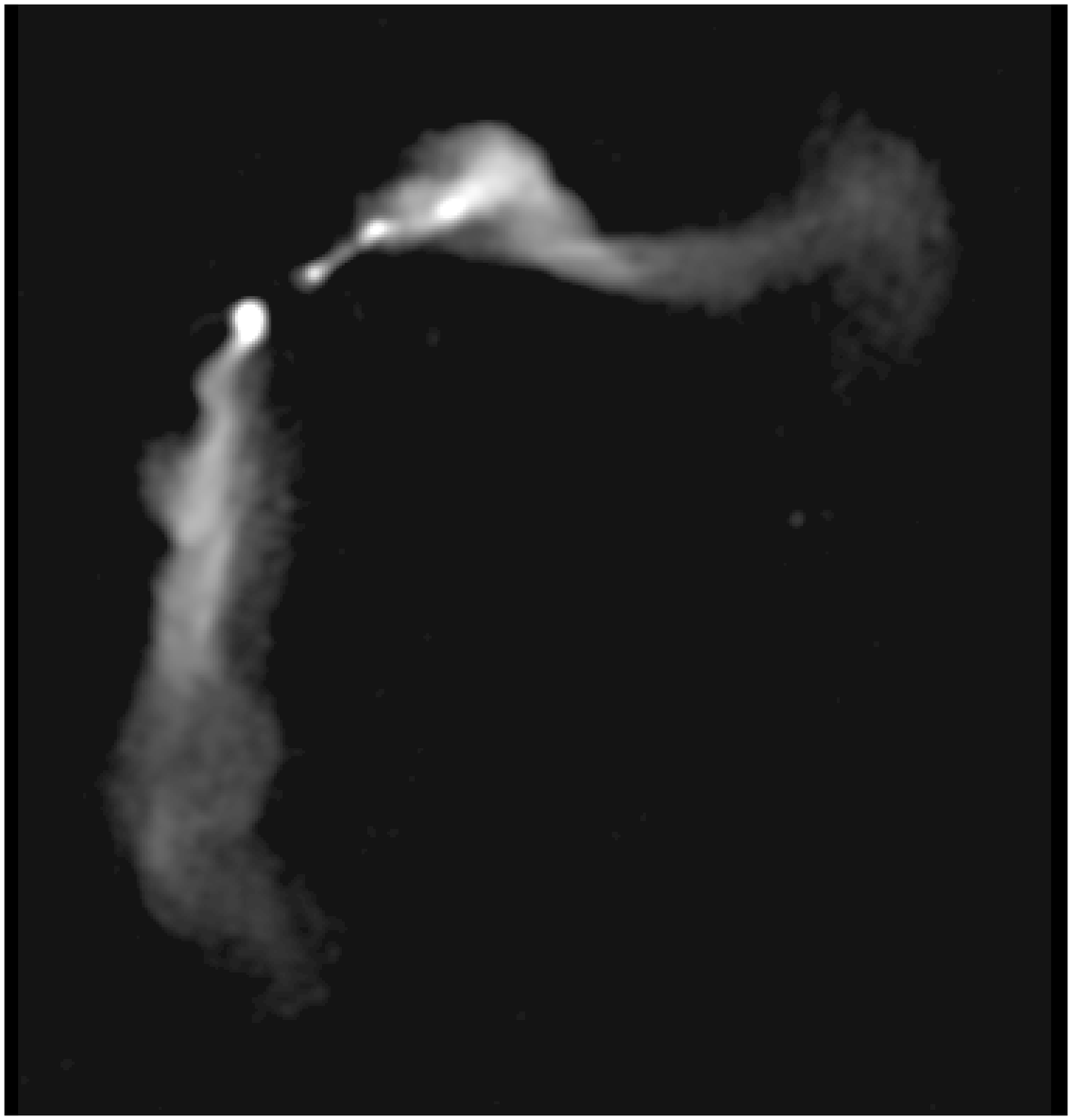}}
\scalebox{0.224}{\includegraphics{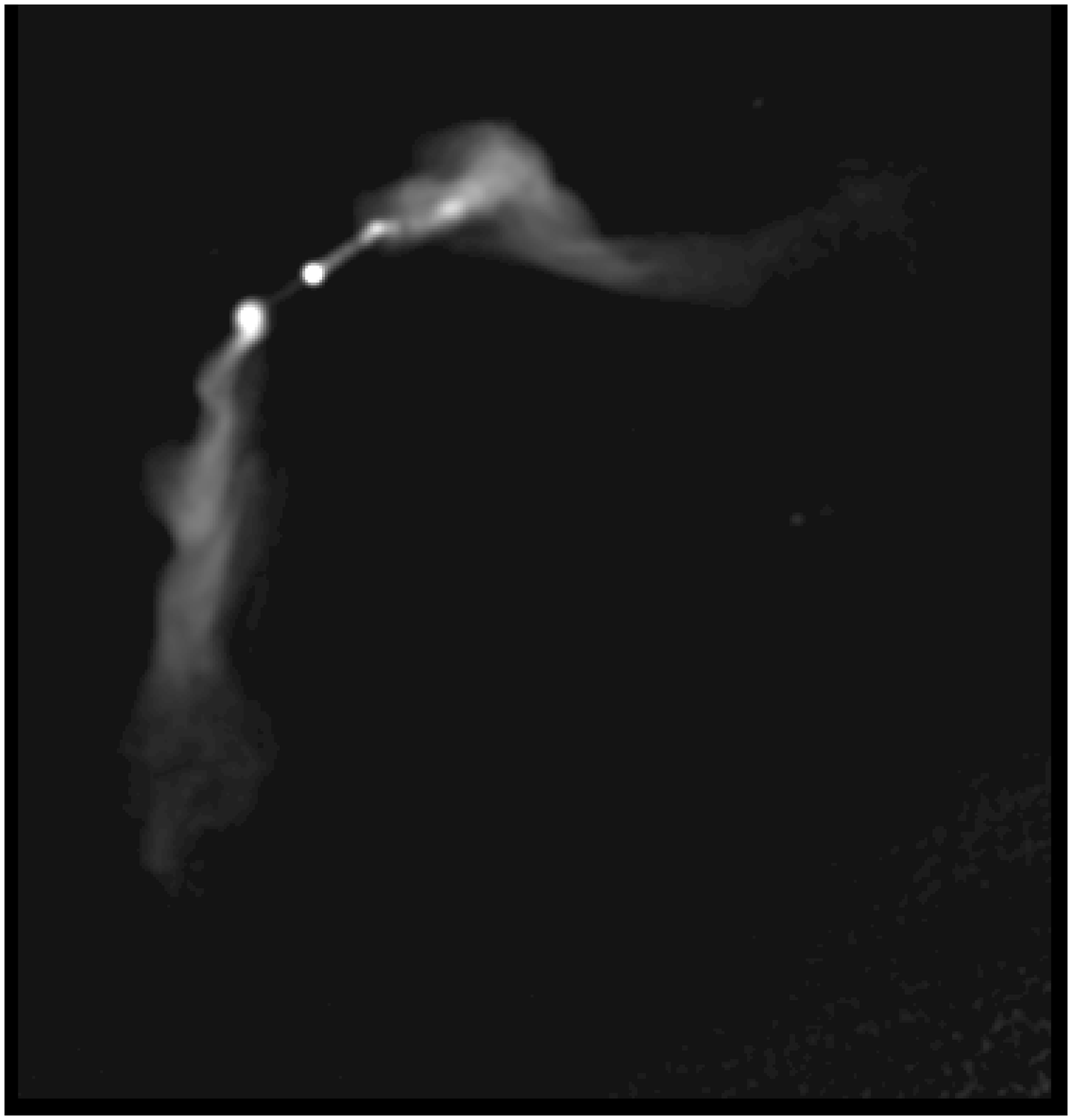}}
\end{center}
\vspace{-2ex}
\caption{Three VLA images of 3C465:  left, 327 MHz;  center, 1.4 GHz;
  right, 4.8 GHz;  from Eilek \& Owen 2002 and references
  therein. This shows the  rapid decay of the radio tails at high
  frequency, that is the steepening of the radio spectrum going out
  along the tail.  This effect is usually ascribed to synchrotron
  losses of the electrons, but this interpretation may be too
  simple.  These images 
  also illustrate how much of a source can be hidden at high
  frequencies; without good low-frequency   images we cannot
  understand the full story  of the radio source.  } 
\end{figure}

Spectral steepening is often interpreted very simply, 
namely, that the high-energy electrons have been depleted by
synchrotron losses (thus the name ``spectral aging'').   The spectrum
steepens at the frequency radiated by electrons whose
radiative lifetime equals the age of the source.  
 If the magnetic field is known (or assumed, say
from equipartition analyses), the age of the radiating plasma can be
derived.  However, the results here are often inconsistent with other
estimates which find a longer age for the radio source ({\it e.g.}
O'Donoghue {\it et al.\,}1993;  Blundell \& Rawlings 2000;  Eilek
2004).

Because source ages are important to many aspects of source modelling,
the simple interpretation of spectral  steepening must be tested, 
not just assumed.  We suggest the best way to do this is to develop
careful dynamic models of some nearby, bright radio sources. Doing this
requires good images over a broad frequency range, {\it including the
lowest possible frequencies}:  in what parts of the source does the
spectrum steepen, and what are the likely plasma dynamics there? Is simple
aging plausible, or must we consider {\it in situ} acceleration or
magnetic field variability, or both?

\subsection{Cautionary Note}

Both of these analyses will be  complicated by inhomogeneities in the
source, especially the magnetic field.  On small scales, we expect the
field to be filamented; on larger scales the dynamics of the source
will determine the magnetic field structure.  We know that magnetic
inhomogeneities can complicate  interpretation of synchrotron spectra
({\it e.g.}, a single electron energy in a power law magnetic field
distribution  can also give a power law synchrotron spectrum;  Eilek
\& Arendt 1996).   In some applications this complication can be dealt
with in terms of mean fields or a two-field approximation (Eilek {\it
  et al.\,}1997);  but 
observations should also be designed with the highest possible
resolution and using nearby sources.

\section{Radio Galaxy Physics}

We have already pointed out that radio galaxies look quite different
at low and high frequencies.  Combining LWA images with higher
frequency images from other telescopes will enable us to probe
the life cycles and internal physics of these sources.  In addition,
the LWA by itself will add to our understanding of these galaxies.
Two examples of the latter are low-frequency absorption studies and
searches for old radio galaxies.

\subsection{Probing the Gas in the Parent Galaxy}

The LWA opens up an exciting new possibility:  studying the interstellar
medium in the gas in the parent galaxy {\it via} detection of
free-free absorption from that gas.  The $\nu^{-2.1}$
behavior of the opacity means the LWA will be able to detect absorbing
clouds that would be invisible at higher frequencies. 
Free-free absorption will be easy to distinguish from a
low-energy cutoff in the electron distribution, by its different radiation
spectrum, as well as by its
  patchy nature (in a well resolved source). 
Most of our knowledge of the interstellar medium (ISM) in an
elliptical galaxy comes from X-rays, which detect the hot phase
(typically $T \sim 10^7$K, $n \sim 0.1$cm$^{-3}$;  Brighenti \& Mathews
1997).  Detection of free-free absorption would allow us to study
quite different, cooler components of the the ISM.  For example,
consider cool gas at $10^4$K in pressure balance with the hot ISM.
Cool clouds in the galactic core would be  just opaque 
 at 50 MHz if they are 1 pc in size;  in the lower-pressure
outer regions of the galaxy, kpc-sized clouds will be just opaque. 
Such clouds would easily be detectable in the  spectrum
of the background source, and possibly in  high-resolution images.

\subsection{Where are the old sources?}

With good low-frequency data we should be able to answer a
longstanding problem: where are the old radio galaxies?  Current
models of radio source physics predict  that the radio spectrum will
steepen as the source ages.  This holds whether or not the jet remains
active,  long as the magnetic field in the source
remains somewhat steady  (Eilek \& Shore 1989).   The models also 
predict that the sources should remain bright at low frequencies,
where synchrotron losses have not yet had an impact.  

These models therefore predict the existence of
a large population of radio sources which
have steep spectra around 1 GHz but are bright at 100 MHz.  However,
such sources are uncommon;  we have not found nearly as many as the
models predict.  Where can they be?  Are the models seriously wrong?
One possibility is that magnetic field dissipates rapidly when the
central engine shuts down.  This could happen, for instance, if the
magnetic field is supported by jet-driven turbulence in the extended
source. If this is the case, then old, steep-spectrum radio
sources would be much fainter than the current models predict.  The
LWA will easily be able to detect these sources. If such sources are
searched for and not found, then the ``last chance'' will be gone, and
will definitely know that our general model of radio galaxies needs
serious revision.

\section{Radio Haloes in Clusters of Galaxies}

An important use of the LWA will be the  direct study of steep-spectrum
synchrotron sources.  Radio haloes in clusters of galaxies are 
one interesting example. The plasma atmosphere in a 
cluster of galaxies contains a significant  nonthermal component, just
as the ISM in our galaxy does.  
The nonthermal component is seen {\it via} its synchrotron
emission.  As illustrated in Figure 2, the synchrotron halo is usually
a diffuse, cluster-wide source,  often obscured by small, bright radio
galaxies within the cluster. 

\begin{figure}[!ht]
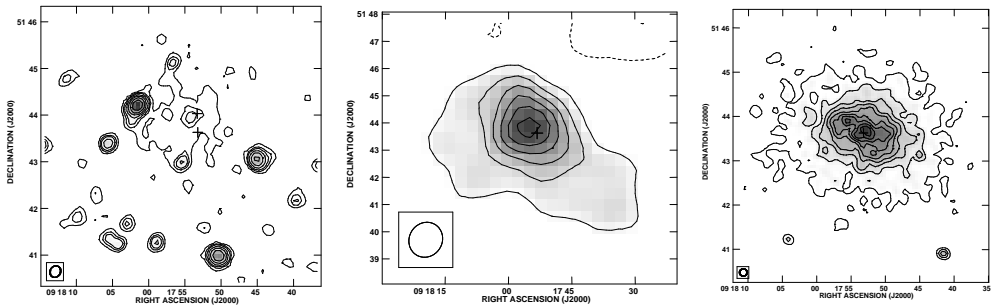

\begin{center}
\scalebox{0.23}{\includegraphics{Eilek1fig2a.ps}}
\scalebox{0.242}{\includegraphics{Eilek1fig2b.ps}}
\scalebox{0.21}{\includegraphics{Eilek1fig2c.ps}}
\end{center}
\vspace{-4ex}
\caption{Radio haloes reveal nonthermal plasma in the intraclutser
  medium, when they can be detected.  Abell 773, from Markovi\'c (2004), is an
  example. Left, a high-resolution 1.4 GHz VLA image showing the
  embedded galactic sources in the cluster.  Center, a low-resolution
  1.4 GHz image of the same cluster,  with the embedded sources removed
  to reveal the diffuse halo.  Right, an archival CHANDRA image of the
  cluster, illustrating that the extent of the radio halo is
  similar to that of the thermal, X-ray loud intracluster medium.} 
\end{figure}

The existence of this nonthermal component
tells us the cluster is not fully relaxed.  Both 
the relativistic particles and the (turbulent) magnetic field must be
re-energized over the life of the cluster.
Until recently it was thought that haloes or relics were rare, and
thus connected to some unusual event, such as a major merger.  We now
have evidence that they are more common (Markovi\'c, 2004) and that
they occur in quiescent 
clusters as well as in actively merging ones.   This suggests
that the dynamic history of clusters is not as simple as has been thought;
there are ongoing energy sources  even in apparently quiescent clusters.  
Fully understanding this synchrotron emission  will tell
us much about the dynamical state and history of the cluster.  

In order to pursue this issue we need good images of radio haloes in
a large number of clusters, both merging and quiescent. 
However, with current instruments and much hard work we are just at
the limits of being able to address this question.  The haloes are
faint, and contaminated by embedded radio galaxies in the cluster.
We need sensitive observations on a wide range of spatial scales to
sort them out.  Because haloes are usually steep spectrum, the LWA
will be both necessary and ideal for this work.

\section{The High Redshift Universe}

There is still much to learn about high-redshift synchrotron sources.
Because they are likely to have steep
radio spectra, high-$z$  sources are best studied --- or simply
detected --- at low frequencies.   We expect them to have steep
spectra because inverse Compton
losses on the microwave background increase with redshift $\propto (1 
+ z)^4$, and because high-frequency spectral breaks are redshifted to
lower frequencies. This is 
consistent with radio galaxies we know about now.  They are generally
the brighter, FRII-type objects, and do tend to have steep 
radio spectra ({\it e.g.} de Breuck {\it et al.\,}2000).  If the nearby
universe can be used as an analogy, FRI-type radio
galaxies and strong starbursts should also exist at high $z$. We
might expect these objects to have weaker magnetic fields than the
FRII's (as they do in the nearby universe), which would shift their
spectral break to even lower frequencies.  We thus expect a large
population of faint, high-$z$ synchrotron sources is waiting to be
studied.  This work calls for the high resolution and high
sensitivity of the LWA. 

One important question in this area is  the demographics of synchrotron 
sources in the early universe.  How abundant are 
star-forming galaxies and massive, accreting black holes at early
epochs?  To answer this question,  we must first detect the objects;
follow-up multiband spectra should enable us to separate the black
holes from the starburst galaxies.  But if the initial radio searches
are carried out only at GHz frequencies, as they must be today, we may
well be missing a large number of the objects.   

Another question is  the physics of radio galaxies at high
redshift.  Are high-$z$ sources just distant  analogues of nearby
sources, subject to stronger Compton losses, or is  the situation more
complex?   FRII sources are not typical of the nearby radio galaxy
population;  tailed, FRI types are much more common.  Will more
sensitive searches, at low frequency, find large numbers of faint,
steep-spectrum FRI sources?  We pointed out above that spectral
steepening is not yet understood in nearby radio sources, particularly
FRI's.  How  will current models survive when tested in the harsher
environment at high redshift?

\section{Closing comments}

In this paper we have  discussed a few of the important
contributions the LWA can make to extragalactic synchrotron physics;
many more good ideas will no doubt appear.  Much of the
science, however, 
depends on getting the highest possible resolution, down to
a few arcseconds.  We thus look forward to creative solutions to the
imaging problems which the ionosphere will impose on us.

\end{document}